\PassOptionsToPackage{nameinlink}{cleveref}
\documentclass[a4paper,USenglish,cleveref,autoref,thm-restate,pdfa]{lipics-v2021}
\hideLIPIcs
\nolinenumbers
\usepackage{cite}
\usepackage[linesnumbered,ruled]{algorithm2e}
\usepackage{booktabs}
\usepackage{tikz}
\usepackage{float}
\usepackage{needspace}
\usetikzlibrary{arrows.meta}
\hypersetup{hypertexnames=false}

\definecolor{defblue}{rgb}{0.121,0.47,0.705}
\NewCommandCopy{\OriginalEmph}{\emph}
\RenewDocumentCommand{\emph}{m}{\textcolor{defblue}{\OriginalEmph{#1}}}
\SetArgSty{OriginalEmph}

\title{Tight Competitive Bounds
for Online~Matching~on~Forests}
\titlerunning{Tight Competitive Bounds for Online Matching on Forests}

\author{Bart{\l}omiej Bosek}
{Institute of Theoretical Computer Science, Faculty of Mathematics and Computer Science, Jagiellonian University, Krak{\'o}w, Poland}
{bartlomiej.bosek@uj.edu.pl}
{https://orcid.org/0000-0001-8756-3663}
{}

\author{Grzegorz Gutowski}
{Institute of Theoretical Computer Science, Faculty of Mathematics and Computer Science, Jagiellonian University, Krak{\'o}w, Poland \and \url{https://grzegorz.gutowscy.pl/}}
{grzegorz.gutowski@uj.edu.pl}
{https://orcid.org/0000-0003-3313-1237}
{}

\author{Paweł Putra}
{Institute of Informatics, University of Warsaw, Poland}
{p.putra@uw.edu.pl}
{https://orcid.org/0009-0006-3703-680X}
{}

\author{Anna Zych-Pawlewicz}
{Institute of Informatics, University of Warsaw, Poland}
{anka@mimuw.edu.pl}
{https://orcid.org/0000-0002-5361-8969}
{}

\authorrunning{B. Bosek, G. Gutowski, P. Putra, and A. Zych-Pawlewicz}
\Copyright{Bart{\l}omiej Bosek, Grzegorz Gutowski, Pawe{\l} Putra, and Anna Zych-Pawlewicz}

\ccsdesc[100]{Mathematics of computing~Discrete mathematics}
\ccsdesc[100]{Theory of computation~Online algorithms}
\keywords{Online matching, fractional matching, fractional vertex cover, trees, competitive analysis}
\funding{This work is a result of research conducted within research project number\\
\mbox{2023/51/B/ST6/02833} financed by the National Science Centre, Poland (Bartłomiej Bosek, Grzegorz Gutowski, Paweł Putra and Anna Zych-Pawlewicz).}

\newcommand{\E}{\mathbb{E}}
\newcommand{\alg}{\mathrm{ALG}}
\newcommand{\opt}{\mathrm{OPT}}

\newcommand{\append}{\mathsf{append}}
\newcommand{\create}{\mathsf{create}}
\newcommand{\set}[1]{\left\{#1\right\}}

\begin{document}
\maketitle

\begin{abstract}
We study the online matching problem, its fractional relaxation, and the fractional online vertex cover problem on forests whose edges arrive one at a time in an arbitrary order.
Matching decisions are irrevocable, and vertex-cover weights can only increase over time.
For online matching, we present a randomized algorithm with expected competitive ratio $4/7$ against an oblivious adversary, improving the previous $5/9$ guarantee.
We also present a deterministic algorithm that simultaneously maintains a $4/7$-competitive fractional matching and a $7/4$-competitive fractional vertex cover.
This deterministic algorithm improves the previous $5/9$ guarantee for fractional matching and the $11/6$ guarantee for fractional vertex cover.
We prove that $4/7$ is optimal for both randomized and fractional matching, even on growing trees of maximum degree four.
\end{abstract}

\section{Introduction}
\label{sec:introduction}

In this paper we consider the matching problem in the online edge
arrival model.
In this model,
the vertex set is known in advance and edges arrive one at a time.
Each edge must be accepted or rejected upon arrival.
The algorithm maintains a set of edges forming a matching, i.e., no two
selected edges share an endpoint.
The objective is to maximize the (expected) size of the matching, and
the challenge is that the decisions of the algorithm are irrevocable,
i.e., once an edge is added to the matching, it must
remain there for the whole runtime of the algorithm.
For this reason, we cannot hope to maintain an offline-optimal matching at every time step.
Instead, we aim for a \emph{competitive ratio} as close to $1$ as
possible, i.e., a guarantee that the expected size of our matching is
at least this fraction of the offline optimum.

A greedy deterministic algorithm guarantees a solution
half the size of a maximum matching, and there is no hope for a
deterministic algorithm to beat
the competitive ratio of $\frac{1}{2}$ even on a path of length three.
Thus, the question of whether one could beat the ratio of $\frac{1}{2}$
using randomization was a central open question in the area
for a long time. Recently Gamlath, Kapralov,
Maggiori, Svensson and Wajc~\cite{GamlathKMSW19} showed
that we cannot hope for that either, even on bipartite graphs.

This motivates studying the edge arrival model on more restricted
classes of graphs. The most widely considered restriction in the
literature is to forests. The pessimistic online bounds for deterministic
algorithms also apply to this class; however, we can achieve better
guarantees for randomized algorithms.
Buchbinder, Segev and Tkach gave a randomized algorithm
with competitive ratio
$\frac{5}{9} \approx 0.5556$~\cite{BuchbinderST17,BuchbinderST19} and proved
that this guarantee is optimal within their
\textsc{Min-Index} framework.
Their upper bound for unrestricted algorithms is approximately
$0.5914$, even on trees of maximum degree three.
We close this gap at $4/7\approx0.5714$.
Our algorithm handles arbitrary edge orders, whereas the matching
upper bound already holds on \emph{growing trees}: after the initial
edge, each arrival attaches a new leaf to the existing tree.

In pursuit of this goal, we study the fractional online
matching and the fractional vertex cover problem on forests in the
edge arrival model.
In the fractional online matching problem, each arriving edge
receives an irrevocable weight from $[0,1]$, so that at any vertex
the weights of adjacent edges sum up to at most one.
The objective is to maximize the sum of the assigned weights.
In the fractional vertex cover problem, vertices receive
weights from $[0,1]$ so that every presented edge is covered, i.e.,
the weights assigned to its endpoints sum up to at least one.
The objective is to minimize the total weight of the vertices,
and the decisions are irrevocable in the sense that we may only
ever increase the weights assigned to the vertices from one step
to the next. For the matching, a competitive ratio $c \leq 1$ means
that
after every edge arrival the weight of the matching is
\emph{at least}
$c$ times the offline optimum, whereas for the vertex cover a
competitive ratio of $c \geq 1$ means that the weight of the
vertex cover is \emph{at most} $c$ times the optimum.

Fractional matching and fractional vertex cover are
linear-programming duals, and their offline optima coincide.
The online requirements are different, however: matching weights
are fixed at
arrival, whereas cover weights must remain nondecreasing as
constraints are revealed.
An offline dual certificate in the analysis of a fractional
online matching
algorithm does not immediately translate
to an online fractional vertex cover algorithm.
Indeed, for fractional
matching on forests,
the result of
Buchbinder, Segev and
Tkach~\cite{BuchbinderST19}
implies a $\frac{5}{9}$-competitive algorithm,
while for fractional vertex cover on forests, Balig\'acs et
al.~\cite{BaligacsBDFGKPZ26} obtain an
$11/6 \approx 1.8333$-competitive algorithm
under edge arrivals. The two algorithms differ substantially. Both
solutions~\cite{BuchbinderST19,BaligacsBDFGKPZ26} use weak duality
in the analysis, but neither constructs
a fractional matching and a fractional vertex cover at the same time
in an online manner.

We improve both of the results above with one online algorithm, 
which constructs both a fractional online matching and a
fractional online vertex cover at the
same time, providing competitive ratios of
$\frac{4}{7}\approx 0.5714$ and
$\frac{7}{4}=1.75$, respectively. 
We show that no algorithm
achieves a constant
competitive
ratio of $c>\frac{4}{7}$
for fractional matching.
This also implies that no
randomized online
matching algorithm
achieves an expected
competitive ratio of
$c > \frac{4}{7}$,
as we discuss in the next
section. Our hard instance
is a growing tree,
implying the same
hardness in the easier
vertex arrival model,
and thus establishing
tight competitive ratios
for the online matching
in the vertex arrival
model on forests as well.

For matching, we also
show how to adapt the
fractional algorithm
to obtain a randomized online algorithm with the same competitive ratio.

\subparagraph*{Our results.} We prove the following guarantees.
\begin{restatable}{theorem}{fractionaltheorem}\label{thm:main}\label{thm:fractional}
In the edge arrival model on forests, there is a deterministic
algorithm that simultaneously maintains a $4/7$-competitive fractional
matching and a $7/4$-competitive fractional vertex cover.
\end{restatable}

\begin{restatable}{theorem}{randomizedtheorem}\label{thm:mainrandom}\label{thm:randomized}
In the edge arrival model on forests, there is a randomized online
matching algorithm with expected
competitive ratio $4/7$ against an oblivious adversary.
\end{restatable}

\begin{restatable}{theorem}{hardnesstheorem}\label{thm:mainlb}
For every constant $c>4/7$ there is a growing tree of maximum degree four on
which no fractional or randomized online matching
algorithm is $c$-competitive.
\end{restatable}
\subparagraph*{Outline.}
Section~\ref{sec:related} summarizes related work.
Section~\ref{sec:preliminaries} defines the problems and the arrival model.
Section~\ref{sec:alg} provides and analyzes the algorithm for both fractional problems.
Section~\ref{sec:randomized} constructs and analyzes the randomized matching
algorithm.
Section~\ref{sec:hardness} proves the matching hardness.

\section{Related work}
\label{sec:related}

The online matching problem dates back to the breakthrough
work of Karp, Vazirani and Vazirani~\cite{KarpVV90} in 1990.
The authors studied online matching in an easier
one-sided \emph{vertex arrival} model on bipartite graphs.
This setting and its generalizations have applications in
ad auctions~\cite{BuchbinderJN07}. In the vertex
arrival model, the vertices arrive one at a time together with all
the edges connecting to the vertices presented so far. The one-sided
version assumes that only the vertices on one side of the bipartition
arrive in an online fashion, while the vertices on the other side are
known up front to the algorithm.
Due to the upper bound of $\frac{1}{2}$ for
the competitive ratio of any deterministic algorithm,
the authors in~\cite{KarpVV90} turn to randomization and show a
widely celebrated \textsc{Ranking} algorithm, which admits
a (tight) competitive ratio of $(1-1/e)\approx 0.6321$. The \textsc{Ranking} algorithm
was further studied in several more works~\cite{Birnbaum,DevanurJK13},
resulting in a relatively
simple primal-dual analysis~\cite{DevanurJK13} that provided the basis
for future work.
Fractional online matching was also
studied in the one-sided vertex arrival model,
resulting in an elegant \emph{water filling} algorithm with a
$(1-1/e)$-competitive ratio~\cite{BuchbinderJN07}.
The water-filling algorithm constructs a feasible dual solution online;
for \textsc{Ranking}, the dual variables are also constructed online,
but feasibility holds in expectation.

Wang and Wong~\cite{WangW15} studied the dual fractional vertex cover problem
under vertex arrivals and observed that it generalizes the ski-rental problem. If one restricts vertex
weights to $\{ 0,1 \}$, thus obtaining the integral online
vertex cover problem, one immediately hits a tight competitive ratio of $2$ for deterministic
algorithms, attained by a simple greedy algorithm~\cite{BoyarFKL17,Boyar}.
Wang and Wong~\cite{WangW15} provide an $e/(e-1)\approx 1.5820$-competitive
algorithm
for the
fractional variant in the one-sided vertex arrival model.
This competitive ratio is the reciprocal of the $(1-\frac{1}{e})$ ratio of the \textsc{Ranking}
algorithm.
Their algorithm constructs the vertex cover and the matching
simultaneously,
and the competitive ratio follows naturally from weak
duality.
More interestingly, the algorithm of Wang and Wong~\cite{WangW15} can
be applied to
the general vertex arrival model, where it achieves a
competitive
ratio of $\frac{1}{\Gamma} \approx 1.9008$ for the fractional online
vertex cover problem, while the best known lower bound is
approximately $1.8244$~\cite{Lu26}. As
a byproduct, from the primal-dual analysis Wang and Wong derive a
competitive ratio of
$\Gamma \approx 0.5261$ for the fractional online matching problem in
the general vertex arrival model, later shown tight by
Tang~\cite{Tang26}.
Subsequently, Gamlath, Kapralov,
Maggiori, Svensson and Wajc~\cite{GamlathKMSW19}
show how to round the weights produced by the algorithm of Wang and
Wong to obtain a randomized algorithm for the online matching
problem in the general
vertex arrival model, achieving a competitive ratio strictly larger
than $\frac{1}{2}$, thus answering a long-standing open problem.

The edge arrival model turns out to be substantially harder
for the algorithm. In
the same paper~\cite{GamlathKMSW19}, it was shown that there is no
algorithm for online fractional matching achieving a competitive
ratio larger than $\frac{1}{2}$ by any constant. Hardness for
fractional online matching immediately implies the same hardness
for randomized online matching: every randomized online matching
induces a fractional one by assigning each edge its probability of being
selected. The converse is more delicate: Buchbinder, Naor
and Wajc~\cite{BuchbinderNW23} study the constraints that make
lossless rounding possible.
Just recently, Tang and Zhang showed
that also for the fractional vertex cover problem under edge arrivals
one
cannot hope for a competitive ratio smaller than $2$~\cite{TangZ26},
even on bipartite graphs. These hardness results make structural
restrictions essential for improving the competitive guarantees under
edge arrivals.

Buchbinder, Segev and Tkach~\cite{BuchbinderST19} established a
$\frac{5}{9} \approx 0.5556$-competitive randomized online
matching algorithm for forests. Taking the probabilities of
accepting the edges as their weights yields a fractional
algorithm with the same guarantee. 
Their \textsc{Min-Index} framework maintains an ordered collection of
matchings. Each arriving edge is added to the first matching that can
accept it, or rejected if none can. The output matching is chosen
according to probabilities fixed in advance. They show that no
instantiation of this framework achieves a competitive ratio greater
than $\frac{5}{9}$ on forests.
For graphs of maximum degree~$d$ known in advance, they also give
a deterministic algorithm that maintains a fractional matching and
a fractional vertex cover simultaneously online, with competitive
ratio $\frac12\bigl(1+\frac{1}{2^d-1}\bigr)$ for matching and its
reciprocal for vertex cover. For $d=3$ this gives
$\frac{4}{7}$ and $\frac{7}{4}$, respectively, but the matching guarantee
tends to $\frac12$ as $d$ grows. 

Buchbinder, Segev and Tkach~\cite{BuchbinderST19} also give a hardness bound 
for fractional matching of $\frac{4}{9-\sqrt{5}} \approx 0.5914$ which holds
already for growing trees with maximum degree three.
Pashkovich and Snow~\cite{PashkovichS25} subsequently attained this
bound for fractional online matching on all graphs of maximum degree
three, thus closing the problem of fractional online matching for
this class of graphs. They also show that no randomized algorithm
achieves a competitive ratio larger than $0.5807$ on graphs of
maximum degree three, and no fractional algorithm achieves a
competitive ratio larger than $0.5889$ on graphs of maximum
degree four. Our results determine the optimal fractional and
randomized ratios for forests without degree restrictions. Moreover,
our upper bound for fractional matching on forests improves the
upper bound
of~\cite{PashkovichS25} within the class of graphs of maximum
degree four, as our construction is in fact a growing tree of
maximum degree four.
For fractional vertex cover on forests under edge arrivals 
\cite{BaligacsBDFGKPZ26} gave an $\frac{11}{6}$-competitive algorithm,
which our algorithm
improves to
$\frac{7}{4}$.

Bromblik~\cite{Bromblik26} established a lower bound of
$(9-\sqrt{5})/4 \approx 1.6910$ for online fractional vertex
cover on trees under edge arrivals.
Together with our $7/4$-competitive algorithm, this leaves
a gap between $(9-\sqrt{5})/4$ and $7/4$ for the optimal
competitive ratio.

In~\cite{BaligacsBDFGKPZ26}
the authors consider the
fractional vertex cover
problem also in the incremental
model, when the entire
edge arrival sequence
is known in advance.
For that model, they
prove an optimal ratio
of $3/2$. Our hardness
bound for fractional
matching applies also
in the
incremental model,
proving that even if
the entire future is
known in advance, the
fractional matching algorithm still
cannot be better than
$\frac{4}{7}$-competitive.

Independently of our work, Bienkowski~\cite{Bienkowski26}
developed an algorithm for forests
under edge arrivals that simultaneously maintains a fractional
vertex cover and a fractional matching, with competitive ratios
$4-\sqrt{5}\approx1.7639$ and
$1/(4-\sqrt{5})\approx0.5669$, respectively.
Our fractional algorithm improves these guarantees to
$7/4=1.75$ and $4/7\approx0.5714$, respectively.

There has
been work on online matching on forests with free disposal, where
one is allowed to discard edges previously accepted to the matching.
In this model, Jiang and Zhang~\cite{JiangZ26} obtained the optimal randomized
ratio $\frac{2}{3}$ on growing trees, and a randomized ratio
$\frac{5}{8}$ on general forests. These guarantees do not apply to
the irrevocable decisions that we consider in this paper.

\section{Preliminaries}\label{sec:preliminaries}

For a graph $G=(V,E)$, a \emph{matching} is a set of edges with disjoint
endpoints, and a \emph{vertex cover} is a set of vertices containing
at least one endpoint of every edge.
A \emph{fractional matching} $\mathbf{x}$ assigns weight~$x_e\in[0,1]$ to each edge~$e$
such that $\sum_{e\ni v}x_e\leq1$ for every vertex $v$;
its value is $\sum_{e\in E}x_e$.
A \emph{fractional vertex cover} $\mathbf{y}$ assigns weight~$y_v\in[0,1]$ to each vertex~$v$
such that every edge $uv$ is \emph{covered}, meaning $y_u+y_v\geq1$;
its cost is $\sum_{v\in V}y_v$.
On forests, all four problems have the same optimum value,
by K\H{o}nig's theorem and weak duality.

We consider randomized algorithms for matching and deterministic algorithms
for fractional matching and fractional vertex cover.

In the online \emph{edge arrival} model, an oblivious adversary fixes
a graph $G=(V,E)$ and an edge order $e_1,\dots,e_m$ in advance.
The algorithm knows $V$ and receives the edges one at a time.
After arrival $t$, it must maintain a feasible solution for
$G_t:=(V,\set{e_1,\dots,e_t})$, using only the revealed prefix
and its own random choices.

Matching decisions are made upon arrival and are irrevocable:
the maintained matching $M_t$ satisfies
$M_{t-1}\subseteq M_t\subseteq M_{t-1}\cup\{e_t\}$, with $M_0=\emptyset$.
A fractional matching algorithm assigns each arriving edge an
irrevocable weight $x_e$.
A fractional vertex-cover algorithm maintains vertex weights $y_v$
that start at zero and can only increase.

The objective is to maximize matching cardinality or fractional matching value,
or to minimize fractional vertex-cover cost.
Write $\alg_t$ for the cardinality, value, or cost of the maintained solution
and $\opt_t$ for the corresponding offline optimum on $G_t$.
An algorithm is $c$-competitive if, for every input and every
$t\in\set{1,\dots,m}$,
\[
\begin{aligned}
\E[\alg_t]&\geq c\cdot\opt_t &&\text{for matching,}\\
\alg_t&\geq c\cdot\opt_t &&\text{for fractional matching,}\\
\alg_t&\leq c\cdot\opt_t &&\text{for fractional vertex cover.}
\end{aligned}
\]
The expectation is over the algorithm's random choices.

\section{Fractional algorithm}\label{sec:algorithm}\label{sec:alg}\label{sec:analysis}\label{sec:verification}

Our algorithm is based on linear programming duality between fractional matching and fractional vertex cover.
Recall that the fractional matching problem on a graph $G=(V,E)$ is the linear program on the left, and the fractional vertex cover problem is its dual, on the right:
\[
\begin{array}{r@{\quad}l}
\max & \sum_{e\in E}x_e\\[2pt]
\text{s.t.} & \sum_{e\ni v}x_e\leq 1\quad\text{for all }v\in V\text{,}\\[2pt]
& x_e\geq 0\quad\text{for all }e\in E\text{,}
\end{array}
\qquad\qquad
\begin{array}{r@{\quad}l}
\min & \sum_{v\in V}y_v\\[2pt]
\text{s.t.} & y_u+y_v\geq 1\quad\text{for all }uv\in E\text{,}\\[2pt]
& y_v\geq 0\quad\text{for all }v\in V\text{.}
\end{array}
\]
Observe that every optimal solution of either linear program assigns weights at most $1$.
By weak duality, the value of every fractional matching of $G$ is at most the cost of every fractional vertex cover of~$G$.

The previous $5/9$-competitive matching algorithm of Buchbinder, Segev and Tkach~\cite{BuchbinderST19} and the $11/6$-competitive vertex cover algorithm of~\cite{BaligacsBDFGKPZ26} use duality to prove their competitive guarantees.
In these analyses, the dual solution serves as a proof certificate; it is not maintained as a second online solution.
The certificates are constructed for each input prefix, and their evolution need not satisfy the online constraints.
For example, in the analysis of the matching algorithm of~\cite{BuchbinderST19}, the dual vertex cover may decrease vertex weights when two trees are joined.
Such decreases violate the online vertex-cover constraints, so this certificate does not yield an online vertex-cover algorithm~\cite{TangZ26}.
Our algorithm uses duality in a stronger, constructive sense: it maintains valid online solutions for both the fractional matching and the fractional vertex cover.
At every input prefix, each solution certifies the competitive guarantee of the other.
Specifically, for a fixed constant $c\in(0,1]$, the maintained fractional matching $\mathbf{x}$ and fractional vertex cover $\mathbf{y}$ satisfy the \emph{dual inequality}
\begin{equation}\label{eq:primal-dual}
\sum_{e\in E(G_t)}x_e\ \geq\ c\cdot\sum_{v\in V}y_v\qquad\text{for every }t\text{.}
\end{equation}
Since $G_t$ is a forest, the matching and vertex-cover linear programs applied to $G_t$ have integral optimal solutions and optimum value $\opt_t$.
Weak duality and~\eqref{eq:primal-dual} give
\[
c\cdot\opt_t\ \leq\ c\cdot\sum_{v\in V}y_v
\ \leq\ \sum_{e\in E(G_t)}x_e\ \leq\ \opt_t\text{.}
\]
Hence the matching is $c$-competitive and the cover is $\frac1c$-competitive.

At any time $t$, let the \emph{load} $m_v$ of a vertex $v$ be the total matching weight assigned to revealed edges incident to $v$, i.e., $m_v:=\sum_{e\in E(G_t):\,e\ni v}x_e$.
The crucial idea behind our approach is to relate the two solutions at every vertex, so that the weight $y_v$ of a vertex $v$ in the cover is a function of its load $m_v$.
For that purpose, we define a \emph{cover function} $\psi\colon[0,1]\to[0,1]$ so that $y_v=\psi(m_v)$.
Under the conditions of \cref{lem:validity}, we define the \emph{match function} $\chi\colon[0,1]^2\to[0,1]$ for $a,b\in[0,1]$ by
\begin{equation}\label{eq:covering-weight}
\chi(a,b):=\min\set{x\in[0,1-\max\{a,b\}]:\psi(a+x)+\psi(b+x)\geq1}\text{.}
\end{equation}
We can then use the cover function $\psi$ and the match function $\chi$ to define \cref{alg:cover-function} -- an online algorithm for fractional matching and vertex cover.

\begin{algorithm}[H]
\caption{Fractional matching and vertex cover defined by a cover function $\psi$}
\label{alg:cover-function}
\KwIn{Vertex set $V$, stream of edges forming a forest $G=(V,E)$, cover function $\psi$ (with match function $\chi$ defined by~\cref{eq:covering-weight})}
\KwOut{A fractional matching and a fractional vertex cover of $G$}
Initialize $m_v\gets 0$ and $y_v\gets 0$ for each $v\in V$\\
\ForEach{online edge $e=(u,v)$}{
  $x_e\gets \chi(m_u,m_v)$\nllabel{line:edge-weight}\\
  $m_u\gets m_u+x_e$, $m_v\gets m_v+x_e$\\
  $y_u\gets \psi(m_u)$, $y_v\gets \psi(m_v)$
}
\Return $\mathbf{x}=(x_e)_{e\in E}$ and $\mathbf{y}=(y_v)_{v\in V}$
\end{algorithm}

The analysis of \cref{alg:cover-function} is organized as follows.
First, in \cref{lem:validity}, we give conditions on the cover function $\psi$ under which the match function $\chi$ is well-defined and the algorithm maintains valid online solutions.
Next, in \cref{lem:competitive}, we give sufficient conditions on the cover function under which the dual inequality in~\cref{eq:primal-dual} holds for \cref{alg:cover-function} and a fixed $c\in(0,1]$.
Finally, we define an explicit cover function $\psi^\star$ and verify that it satisfies the conditions.
Applying both lemmas, we obtain the competitive guarantees.

\begin{lemma}\label{lem:validity}
Suppose that $\psi\colon[0,1]\to[0,1]$ is continuous and nondecreasing, with $\psi(0)=0$ and $\psi(1)=1$.
Then \cref{alg:cover-function} maintains a valid online fractional matching and a valid online fractional vertex cover after every edge arrival.
\end{lemma}
\begin{proof}
Fix any $a,b\in[0,1]$.
Put $h=1-\max\{a,b\}$.
Then $h\in[0,1]$, $a+h,b+h\in[0,1]$, and at least one of $a+h,b+h$ equals $1$.
Since $\psi(1)=1$ and $\psi$ is nonnegative, $x=h$ satisfies all constraints in~\cref{eq:covering-weight}, so the feasible set is nonempty.
The function $x\mapsto\psi(a+x)+\psi(b+x)$ is continuous and nondecreasing on $[0,h]$.
Thus the feasible set is a nonempty closed subset of the compact interval $[0,h]$, and the minimum in~\cref{eq:covering-weight} is attained.
Thus, the value $\chi(a,b)$ is well-defined.

The initialization and update rules in \cref{alg:cover-function} ensure that $m_v=\sum_{e\in E(G_t):\,e\ni v}x_e$ and $y_v=\psi(m_v)$ for every vertex $v$ after $t$ arrivals.
The value of $x_e$ for an edge $e$ never changes once it is assigned when $e$ arrives.

All loads are initially zero.
For $\mathbf{x}$ to be a valid online fractional matching, it remains to prove that all loads remain in $[0,1]$ after every arrival.
For an arriving edge $e=(u,v)$, let $a=m_u$ and $b=m_v$ be its old endpoint loads.
For $a,b\in[0,1]$, the definition of $\chi$ guarantees that $x_e=\chi(a,b)$ satisfies $x_e\geq0$, $a+x_e\leq1$, and $b+x_e\leq1$.
The algorithm increases the endpoint loads to $a+x_e,b+x_e$ and leaves all other loads unchanged.
Thus all loads remain in $[0,1]$, and $\mathbf{x}$ remains a feasible solution.

The initial graph has no edges, so the initial vertex weights are zero and $\mathbf{y}$ is a valid online fractional vertex cover.
For the same edge $e=(u,v)$, the updated endpoint weights sum to $\psi(a+x_e)+\psi(b+x_e)$, which by the definition of $\chi$ is at least $1$, so the new edge is covered.
Since loads only increase and $\psi$ is nondecreasing, vertex weights never decrease, and all earlier edges remain covered.
Hence $\mathbf{y}$ is feasible after every arrival and satisfies the online constraints.
\end{proof}

Now, we give sufficient conditions on the cover function $\psi$ that ensure that the dual inequality in~\cref{eq:primal-dual} holds for a fixed $c\in(0,1]$.
At any time $t$, let $T$ be a component of the forest $G_t$.
We define the \emph{surplus} of $T$ by
\[
S(T):=\sum_{e\in E(T)}x_e-c\cdot\sum_{v\in V(T)}y_v\text{.}
\]
In order to maintain the dual inequality in~\cref{eq:primal-dual}, it would suffice to show that the surplus of any component is nonnegative.
The surplus of a one-vertex component is $0$.
For other components, we fix a constant $s\geq 0$ and maintain a surplus of at least $s$ in every component containing at least one edge.
An arriving edge in a forest joins two distinct components.
We call it an \emph{isolated edge} if both components are isolated vertices, a \emph{leaf edge} if exactly one is an isolated vertex, and a \emph{join edge} otherwise.
Intuitively, adding an isolated edge must create surplus at least $s$, adding a leaf edge must preserve it, and adding a join edge can decrease the combined surplus by at most $s$.

\begin{lemma}\label{lem:competitive}
Let $\psi$ satisfy the assumptions of \cref{lem:validity}, and let $0<c\leq 1$ and $s\geq 0$.
Suppose that the weight $x_{e}$ chosen by \cref{alg:cover-function} for edge $e=(u,v)$ with loads $m_u,m_v$ before the arrival satisfies
\begin{equation}\label{eq:pay}
x_{e}\geq
\begin{cases}
c+s, & \text{for an isolated edge,}\\[2pt]
c\cdot\bigl(1-\psi(m_u)\bigr), & \text{for a leaf edge with new vertex }v\text{,}\\[2pt]
c\cdot\max\{0,1-\psi(m_u)-\psi(m_v)\}-s, & \text{for a join edge.}
\end{cases}
\end{equation}
Then every component $T$ containing an edge satisfies $S(T)\geq s$ at all times.
\end{lemma}
In particular, when the conditions of \cref{lem:competitive} are met, the algorithm maintains the dual inequality in~\cref{eq:primal-dual} for this choice of $c$ and $s$ and is therefore $c$-competitive for fractional matching and $\frac1c$-competitive for fractional vertex cover.
\begin{proof}
We prove the bound by induction on the number of edges that have arrived.
Initially every component is an isolated vertex and has surplus zero.

An arriving edge $e=(u,v)$ joins two distinct components $T_u,T_v$ into a component $T$.
Let $a=m_u$, $b=m_v$ and put $\delta=\max\{0,1-\psi(a)-\psi(b)\}$.
If $\psi(a)+\psi(b)\geq1$, then $x_e=\chi(a,b)=0$.
Otherwise, continuity of $\psi$ and the minimality of $x_e=\chi(a,b)$ give $\psi(a+x_e)+\psi(b+x_e)=1$.
Thus the total cover weight increases by exactly $\delta$, while the matching weight increases by $x_e$, so
\[
S(T)=S(T_u)+S(T_v)+x_e-c\cdot\delta\text{.}
\]

For an isolated edge, both old components have surplus zero and $a=b=0$, so $\delta=1$.
The first bound in~\cref{eq:pay} gives $x_e\geq c+s$, and therefore
$
S(T)\geq0+0+(c+s)-c=s
$.

For a leaf edge, take $v$ to be the new vertex, so $b=0$ and $\delta=1-\psi(a)$.
By the induction hypothesis, $S(T_u)\geq s$, while $S(T_v)=0$.
The second bound in~\cref{eq:pay} gives $x_e\geq c\cdot\delta$, so
$
S(T)\geq s+0+c\cdot\delta-c\cdot\delta=s
$.

For a join edge, both old components have surplus at least $s$ by the induction hypothesis.
The third bound in~\cref{eq:pay} gives $x_e\geq c\cdot\delta-s$, so
$
S(T)\geq2s+(c\cdot\delta-s)-c\cdot\delta=s
$.

All other components retain their surpluses, completing the induction.
\end{proof}

\begin{definition}\label{def:y-final}
The \emph{cover function} $\psi^\star\colon[0,1]\to[0,1]$ is
\begin{equation}\label{eq:y-final}
\psi^\star(m)=
\begin{cases}
0, & 0\leq m\leq\frac17\text{,}\\[2pt]
\frac78m-\frac18, & \frac17\leq m\leq\frac37\text{,}\\[2pt]
\frac76m-\frac14, & \frac37\leq m\leq\frac67\text{,}\\[2pt]
\frac74m-\frac34, & \frac67\leq m\leq1\text{.}
\end{cases}
\end{equation}
\end{definition}

\begin{figure}[htbp]
\centering
\begin{subfigure}[t]{0.48\linewidth}
\centering
\begin{tikzpicture}[baseline=0pt,x=4.8cm,y=3.4cm,>={Stealth[length=2mm]}]
\tikzset{slope label/.style={midway,anchor=center,
  xshift={-12pt*sin(atan((#1)*3.4/4.8))},
  yshift={12pt*cos(atan((#1)*3.4/4.8))}}}
\draw[->,gray] (0,0) -- (1.1,0) node[right,black] {$m$};
\draw[->,gray] (0,0) -- (0,1.13) node[above,black] {$\psi^\star(m)$};
\foreach \a/\lab in {0.142857/{\frac17},0.428571/{\frac37},0.857143/{\frac67},1/{1}}
  \draw (\a,0.012) -- (\a,-0.012) node[below] {$\lab$};
\foreach \a/\lab in {0.25/{\frac14},0.5/{\frac12},0.75/{\frac34},1/{1}}
  \draw (0.008,\a) -- (-0.008,\a) node[left] {$\lab$};
\draw[gray,densely dotted] (0.428571,0) -- (0.428571,0.25) -- (0,0.25);
\draw[gray,densely dotted] (0.857143,0) -- (0.857143,0.75) -- (0,0.75);
\draw[gray,densely dotted] (1,0) -- (1,1) -- (0,1);
\draw[semithick] (0,0) -- ({1/7},0)
  -- node[slope label=7/8] {$\frac78$} ({3/7},{1/4})
  -- node[slope label=7/6] {$\frac76$} ({6/7},{3/4})
  -- node[slope label=7/4] {$\frac74$} (1,1);
\foreach \p in {(0,0),({1/7},0),({3/7},{1/4}),({6/7},{3/4}),(1,1)}
  \fill \p circle[radius=1.1pt];
\end{tikzpicture}
\caption{Cover function $\psi^\star$.}
\label{fig:cover-function-plot}
\end{subfigure}\hfill
\begin{subfigure}[t]{0.48\linewidth}
\centering
\begin{tikzpicture}[baseline=0pt,x=4.8cm,y=3.4cm,>={Stealth[length=2mm]}]
\draw[->,gray] (0,0) -- (1.1,0) node[right,black] {$d$};
\draw[->,gray] (0,0) -- (0,1.13) node[above,black] {$\mu(d),\,\mu'(d)$};
\foreach \a/\lab in {0/{0},0.428571/{\frac37},0.857143/{\frac67},1/{1}}
  \draw (\a,0.012) -- (\a,-0.012) node[below] {$\lab$};
\foreach \a/\lab in {0.142857/{\frac17},0.428571/{\frac37},0.642857/{\frac9{14}},0.857143/{\frac67},1/{1}}
  \draw (0.008,\a) -- (-0.008,\a) node[left] {$\lab$};
\draw[gray,densely dotted] ({3/7},0) -- ({3/7},1.04);
\draw[gray,densely dotted] ({6/7},0) -- ({6/7},1.04);
\draw[gray,densely dotted] (0,{6/7}) -- ({3/7},{6/7});
\draw[gray,densely dotted] (0,{3/7}) -- ({3/7},{3/7});
\draw[gray,densely dotted] (0,{1/7}) -- ({6/7},{1/7});
\draw[gray,densely dotted] (0,1) -- ({6/7},1);
\draw[semithick] (0,{9/14}) -- ({3/7},{6/7}) -- ({6/7},1) -- (1,1);
\draw[semithick] (0,{9/14}) -- ({3/7},{3/7}) -- ({6/7},{1/7}) -- (1,0);
\foreach \p in {(0,{9/14}),({3/7},{6/7}),({3/7},{3/7}),({6/7},1),({6/7},{1/7})}
  \fill \p circle[radius=1.1pt];
\node[below=6pt,fill=white,inner sep=1.5pt] at ({9/14},{13/14}) {$\mu'(d)$};
\node[above=10pt,fill=white,inner sep=1.5pt] at ({9/14},{2/7}) {$\mu(d)$};
\end{tikzpicture}
\caption{Load update functions $\mu$ and $\mu'$.}
\label{fig:updated-loads}
\end{subfigure}
\caption{Functions defining the fractional algorithm.}
\label{fig:cover-function}
\end{figure}
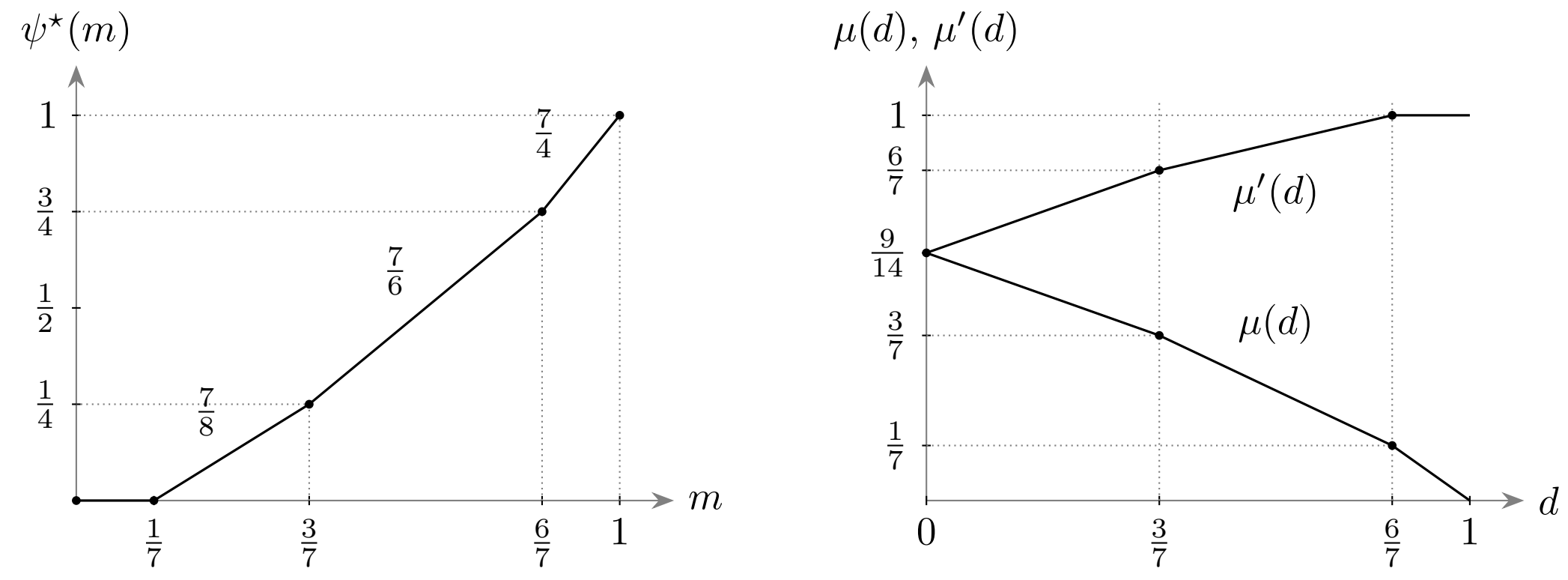

\Cref{fig:cover-function-plot} shows the function.
Observe that the four pieces of $\psi^\star$ are linear, with slopes $0$, $7/8$, $7/6$, and $7/4$.
The pieces agree at their common endpoints and the function is continuous.
It is also nondecreasing and convex, with $\psi^\star(0)=0$ and $\psi^\star(1)=1$.
By \cref{lem:validity}, \cref{alg:cover-function} using $\psi^\star$ maintains valid online solutions for both problems.
We use $\chi^\star$ to denote the match function associated with $\psi^\star$ by~\cref{eq:covering-weight}.
It remains to verify that \cref{lem:competitive} also applies to $\psi^\star$.

\begin{lemma}\label{lem:verify}
For the cover function $\psi^\star$, the weights chosen by \cref{alg:cover-function} satisfy~\cref{eq:pay} with $c=\frac47$ and $s=\frac1{14}$.
\end{lemma}
\begin{proof}
We need to verify the three bounds in~\cref{eq:pay} for isolated, leaf and join edges, with $c=\frac47$ and $s=\frac1{14}$.
Suppose that an edge $e=(u,v)$ arrives.
Let $a=m_u$ and $b=m_v$ be its endpoint loads before the arrival, and let $x_e=\chi^\star(a,b)$ be the weight assigned to~$e$.
By symmetry, we may assume $a\geq b$.

We first express $x_e$ in terms of $a$ and $b$.
If $\psi^\star(a)+\psi^\star(b)\geq1$, then the edge is already covered and $x_e=0$.
This is impossible for an isolated edge, where $a=b=0$.
For a leaf edge, $b=0$ and $\psi^\star(a)=1$, so the required lower bound in~\cref{eq:pay} is $0$; for a join edge, it is $-s\leq0$.
We may therefore assume $\psi^\star(a)+\psi^\star(b)<1$.
By continuity of $\psi^\star$ and the minimality of $x_e$, the updated cover weights sum to exactly one:
\[
\psi^\star(a+x_e)+\psi^\star(b+x_e)=1\text{.}
\]
Both loads increase by $x_e$, so their difference $d=a-b$ is unchanged.
The updated loads therefore depend only on $d$: they are the unique pair with difference $d$ whose cover weights sum to one.
Uniqueness follows because $\psi^\star$ is strictly increasing wherever it is positive.
Define the \emph{load update functions} $\mu,\mu'\colon[0,1]\to[0,1]$ to be uniquely determined functions satisfying $\mu'(d)-\mu(d)=d$, and $\psi^\star(\mu(d))+\psi^\star(\mu'(d))=1$ for every $d\in[0,1]$.
\Cref{fig:updated-loads} gives the plots of these functions.
The functions are continuous and piecewise linear, with changes of slope at $d=\frac37$ and $d=\frac67$.
The covering equality and~\cref{eq:y-final} give the explicit formulas
\begin{equation}\label{eq:load-update-functions}
\mu(d)=
\begin{cases}
\frac9{14}-\frac d2, & 0\leq d\leq\frac37\text{,}\\[2pt]
\frac57-\frac{2d}{3}, & \frac37\leq d\leq\frac67\text{,}\\[2pt]
1-d, & \frac67\leq d\leq1\text{,}
\end{cases}
\qquad
\mu'(d)=
\begin{cases}
\frac9{14}+\frac d2, & 0\leq d\leq\frac37\text{,}\\[2pt]
\frac57+\frac d3, & \frac37\leq d\leq\frac67\text{,}\\[2pt]
1, & \frac67\leq d\leq1\text{.}
\end{cases}
\end{equation}

We have $b+x_e=\mu(d)$ and $a+x_e=\mu'(d)$, so $x_e=\mu(d)-b$.
The three cases in~\cref{eq:pay} depend on whether both, exactly one,
or neither endpoint is isolated before the arrival:
\begin{itemize}
\item
For an isolated edge, $a=b=0$, so $d=0$ and $x_e=\mu(0)-0=\mu(0)=\frac9{14}=c+s$.

\item 
For a leaf edge, $b=0$, so $d=a$ and $x_e=\mu(a)$.
For $a\leq\frac17$, we have $\mu(a)\geq c$ and $\psi^\star(a)=0$, so the bound $\mu(a)\geq c\cdot(1-\psi^\star(a))$ in~\cref{eq:pay} holds.
For $a\geq\frac17$, comparing $\mu$ with~\cref{eq:y-final} gives $\mu(a)=c\cdot(1-\psi^\star(a))$.

\item
For a join edge, observe as above that
\[
\mu(a)\geq c\cdot(1-\psi^\star(a))
\qquad\text{and}\qquad
\mu(b)\geq c\cdot(1-\psi^\star(b))\text{.}
\]
The slopes of $\mu$ are $-\frac12$, $-\frac23$, and $-1$, so increasing its argument by $b$ decreases its value by at least $\frac b2$.
Applying this from $d$ to $a=d+b$ and from $0$ to $b$, respectively, yields
\[
\mu(a)\leq\mu(d)-\frac b2
\qquad\text{and}\qquad
\mu(b)\leq\mu(0)-\frac b2\text{.}
\]
Combining these bounds and using $\mu(d)-b=x_e$ and $\mu(0)=c+s$, we obtain
\begin{align*}
c\cdot\bigl(2-\psi^\star(a)-\psi^\star(b)\bigr)
&\leq \mu(a)+\mu(b)
\leq \mu(d)+\mu(0)-b
=x_e+c+s\text{.}
\end{align*}
Rearranging yields $x_e\geq c\cdot(1-\psi^\star(a)-\psi^\star(b))-s$, as required.\qedhere
\end{itemize}
\end{proof}

We are now ready to restate and prove the main result for the fractional algorithm.
\fractionaltheorem*
\begin{proof}
Consider \cref{alg:cover-function} defined by the cover function $\psi^\star$.
The function $\psi^\star$ satisfies the assumptions of \cref{lem:validity}, as checked after its definition.
\Cref{lem:verify} verifies the payment conditions in~\cref{eq:pay} for $\psi^\star$ with $c=\frac47$ and $s=\frac1{14}$.
\Cref{lem:competitive} therefore gives surplus at least $s$ in every component containing an edge, while isolated vertices have surplus zero.
Summing over components yields~\cref{eq:primal-dual}, and hence both competitive guarantees.
\end{proof}

\section{Randomized algorithm}\label{sec:randomized}

We now use the cover function $\psi^\star$ and its match function $\chi^\star$ to obtain a randomized online matching algorithm.
For the analysis, we adapt the rounding argument of Jiang and Zhang~\cite{JiangZ26}.
Throughout this section, put $c=\frac47$ and $s=\frac1{14}$, as in~\cref{lem:verify}.
The algorithm maintains a matching $M$ and deterministic loads $m_v$, which will equal the probabilities that the vertices are matched.
An arriving edge $e=(u,v)$ joins two distinct components of the forest.
Since their random choices are independent and the old endpoint loads are $a=m_u$ and $b=m_v$, both endpoints are unmatched with probability $(1-a)(1-b)$.
To accept $e$ with probability $x_e$, we must have $x_e\leq(1-a)(1-b)$.
The fractional weight $\chi^\star(a,b)$ can exceed this bound: for example,
\[
\chi^\star\bigl(\tfrac{13}{14},\tfrac1{14}\bigr)=\tfrac1{14}
>\bigl(1-\tfrac{13}{14}\bigr)\bigl(1-\tfrac1{14}\bigr)=\tfrac{13}{196}\text{.}
\]
We therefore define the \emph{capped match function} by
\begin{equation}\label{eq:capped-weight}
  \check{\chi}(a,b)=\min\{\chi^\star(a,b),(1-a)(1-b)\}\text{,}
\end{equation}
and we assign $x_e=\check{\chi}(a,b)$ to the edge.
When both endpoints are unmatched, we accept the edge with probability $x_e/((1-a)(1-b))$.
The loads increase by $x_e$ regardless of whether the edge is accepted.

\begin{algorithm}[H]
\caption{Randomized matching}
\label{alg:randomized}
\KwIn{Vertex set $V$, stream of edges forming a forest $G=(V,E)$}
\KwOut{A matching of $G$}
Initialize $M\gets\emptyset$ and $m_v\gets0$ for each $v\in V$\\
\ForEach{online edge $e=(u,v)$}{
  $a\gets m_u$, $b\gets m_v$\\
  $p\gets(1-a)(1-b)$, $x_e\gets\check{\chi}(a,b)$\\
  \If{$p>0$ and neither $u$ nor $v$ is matched by $M$}{
    With probability $x_e/p$, add $e$ to $M$\\
  }
  $m_u\gets a+x_e$, $m_v\gets b+x_e$
}
\Return $M$
\end{algorithm}

The capped edge weights depend only on the revealed input, so the loads, which are sums of incident edge weights, are deterministic.
We first prove that each load equals the probability that the corresponding vertex is matched.
We then show that the capped weights satisfy the payment bounds in~\cref{eq:pay} and use these bounds to maintain a surplus against every offline matching.

\begin{lemma}\label{lem:marginals}
\Cref{alg:randomized} accepts each arriving edge $e$ with probability $x_e$.
\end{lemma}
\begin{proof}
All loads start at zero.
For endpoint loads $a,b\in[0,1]$, we have
\[
0\leq x_e\leq(1-a)(1-b)\leq\min\{1-a,1-b\}\text{,}
\]
so the updated loads $a+x_e$ and $b+x_e$ also lie in $[0,1]$.
Assign an independent random variable to every arriving edge.
The matching in a component depends only on the variables assigned to its own edges.
Distinct components use disjoint sets of these variables, so their matchings are independent.

We prove by induction that each load $m_v$ equals the probability that $v$ is matched.
Initially all vertices are unmatched and all loads are zero.
Suppose that $e=(u,v)$ arrives with old endpoint loads $a,b$.
Its endpoints lie in distinct components, so by independence and the induction hypothesis they are both unmatched with probability $p=(1-a)(1-b)$.
If $p>0$, the probability of accepting $e$ is $p\cdot(x_e/p)=x_e$.
If $p=0$, then $x_e=0$ and the edge is rejected.
Since $e$ is accepted only when both endpoints are unmatched, their matching probabilities increase by $x_e$, to $a+x_e$ and $b+x_e$, respectively.
These are exactly the updated loads; all other loads and matching probabilities remain unchanged.\qedhere
\end{proof}

We now show that capping preserves the payment bounds in~\cref{eq:pay}.
For isolated and leaf edges, the cap does not change the assigned weight.
For join edges, it suffices to show that the cap $(1-a)(1-b)$ is itself large enough to satisfy the required payment bound.
The following technical lemma gives the estimate needed for this step.

\begin{lemma}\label{lem:free-probability}
For all $0\leq b\leq a\leq1$,
\[
(1-a)(1-b)\geq c\cdot\bigl(1-\psi^\star(a)-\psi^\star(b)\bigr)-s\text{.}
\]
\end{lemma}
\begin{proof}
Since $c-s=\frac12$, it suffices to show that
\[
F(a,b):=(1-a)(1-b)+c\cdot\psi^\star(a)+c\cdot\psi^\star(b)\geq\frac12\text{.}
\]
Convexity and the linear pieces in~\cref{eq:y-final} give the two bounds
\[
c\cdot\psi^\star(m)\geq\frac m2-\frac1{14}
\qquad\text{and}\qquad
c\cdot\psi^\star(m)\geq m-\frac37
\qquad\text{for every }m\in[0,1]\text{.}
\]
If $a\leq\frac12$, apply the first bound at both $a$ and $b$ to obtain
\[
\begin{aligned}
F(a,b)
&\geq (1-a)(1-b)+\left(\frac a2-\frac1{14}\right)+\left(\frac b2-\frac1{14}\right)\\
&=\frac67-\frac{a+b}{2}+ab
=\frac{17}{28}+\left(\frac12-a\right)\left(\frac12-b\right)
\geq\frac{17}{28}\geq\frac12\text{.}
\end{aligned}
\]
If $a\geq\frac12$, apply the second bound at $a$ and the first at $b$ to obtain
\[
\begin{aligned}[b]
F(a,b)
&\geq (1-a)(1-b)+\left(a-\frac37\right)+\left(\frac b2-\frac1{14}\right)\\
&=\frac12-\frac b2+ab
=\frac12+b\left(a-\frac12\right)
\geq\frac12\text{.}
\end{aligned}\qedhere
\]
\end{proof}

\Cref{lem:verify} establishes the payment bounds in~\cref{eq:pay} for~\cref{alg:cover-function} using $\psi^\star$.
Its proof gives these bounds for arbitrary endpoint loads.
We now show that capping preserves them, yielding the analogous guarantee for~\cref{alg:randomized}.

\begin{lemma}\label{lem:capped-payments}
For the cover function $\psi^\star$, the weights chosen by \cref{alg:randomized} satisfy~\cref{eq:pay} with $c=\frac47$ and $s=\frac1{14}$.
\end{lemma}
\begin{proof}
We need to verify the three bounds in~\cref{eq:pay} for isolated, leaf and join edges, with $c=\frac47$ and $s=\frac1{14}$.
\begin{itemize}
\item
For an isolated edge, $a=b=0$ and the cap is $1$, so $\check{\chi}(0,0)=\chi^\star(0,0)=\mu(0)=c+s$.
\item
For a leaf edge, take $b=0$.
The cap is $1-a$, and $\chi^\star(a,0)\leq1-a$ by the definition of the match function.
Hence $\check{\chi}(a,0)=\chi^\star(a,0)=\mu(a)\geq c\cdot(1-\psi^\star(a))$.
\item
For a join edge, put $\delta=\max\{0,1-\psi^\star(a)-\psi^\star(b)\}$.
The pointwise join bound gives $\chi^\star(a,b)\geq c\cdot\delta-s$.
By~\cref{lem:free-probability} and nonnegativity, the cap also satisfies $(1-a)(1-b)\geq c\cdot\delta-s$.
Their minimum therefore satisfies $\check{\chi}(a,b)\geq c\cdot\delta-s$.\qedhere
\end{itemize}
\end{proof}

As before, put $y_v=\psi^\star(m_v)$.
Capping can leave an edge with $y_u+y_v<1$, so these weights need not form a vertex cover.
We instead compare with an arbitrary matching $\mathcal{M}$ of a component $T$.
In the vertex-weight sum defining $S(T)$, we replace the total weight of the endpoints of each edge of $\mathcal{M}$ by $1$, leaving the weights of unmatched vertices unchanged.
Let $V(\mathcal{M})$ be the set of vertices matched by $\mathcal{M}$, and define the \emph{matching surplus}
\[
S(T,\mathcal{M}):=\sum_{e\in E(T)}x_e
-c\cdot\left(|\mathcal{M}|+\sum_{v\in V(T)\setminus V(\mathcal{M})}y_v\right)\text{.}
\]
For the empty matching $\mathcal{M}=\emptyset$, $S(T,\emptyset)$ is the surplus $S(T)$ from~\cref{sec:alg}.
For a maximum matching, nonnegative matching surplus implies $\sum_{e\in E(T)}x_e\geq c\cdot\opt(T)$, where $\opt(T)$ is the size of a maximum matching of $T$.
The next lemma shows that~\cref{alg:randomized} maintains matching surplus at least $s$.

\begin{lemma}\label{lem:randomized-certificate}
After every arrival, \cref{alg:randomized} maintains $S(T,\mathcal{M})\geq s=\frac1{14}$ for every component $T$ containing an edge and every matching $\mathcal{M}$ of $T$.
\end{lemma}
\begin{proof}
We proceed by induction on the arrivals.
An isolated vertex has load zero and matching surplus zero for its only matching, the empty matching.
Suppose that $e=(u,v)$ joins components $T_u,T_v$ into $T$.
Let $a=m_u$, $b=m_v$ be the old loads and put
\[
\delta:=\max\{0,1-\psi^\star(a)-\psi^\star(b)\}\text{.}
\]
The uncapped update $\chi^\star(a,b)$ would increase $y_u+y_v$ by exactly $\delta$, as in the proof of~\cref{lem:competitive}.
By~\cref{eq:capped-weight}, $x_e=\check{\chi}(a,b)\leq\chi^\star(a,b)$.
Since $\psi^\star$ is nondecreasing, the actual increases $\Delta y_u,\Delta y_v$ satisfy
\begin{equation}\label{eq:cover-increase}
\Delta y_u\geq0\text{,}\qquad \Delta y_v\geq0\text{,}\qquad
\Delta y_u+\Delta y_v\leq\delta\text{.}
\end{equation}

Fix a matching $\mathcal{M}$ of $T$, and let $\mathcal{M}_u,\mathcal{M}_v$ be its restrictions to the old components.
We claim that
\begin{equation}\label{eq:matching-surplus-merge}
S(T,\mathcal{M})\geq S(T_u,\mathcal{M}_u)+S(T_v,\mathcal{M}_v)+x_e-c\cdot\delta\text{,}
\end{equation}
where the surpluses on the right are taken before the arrival.
If $e\in \mathcal{M}$, then $u,v$ are unmatched in $\mathcal{M}_u,\mathcal{M}_v$.
Their old cover weights disappear from the sum over unmatched vertices, and the new edge contributes one to $|\mathcal{M}|$.
Thus the change in surplus is exactly
\[
x_e-c\cdot\bigl(1-\psi^\star(a)-\psi^\star(b)\bigr)
\geq x_e-c\cdot\delta\text{.}
\]
If $e\notin \mathcal{M}$, the size of the comparison matching is unchanged, while the total edge weight increases by $x_e$.
The sum of weights of unmatched vertices increases by at most $\Delta y_u+\Delta y_v\leq\delta$, by~\cref{eq:cover-increase}.
The change in surplus is therefore at least $x_e-c\cdot\delta$, proving~\cref{eq:matching-surplus-merge} in this case as well.

We now apply the bounds from~\cref{lem:capped-payments}, with the same accounting as in~\cref{lem:competitive}.
For an isolated edge, the old surpluses are zero, $\delta=1$, and $x_e\geq c+s$.
For a leaf edge, the old surpluses sum to at least $s$ and $x_e\geq c\cdot\delta$.
For a join edge, the old surpluses sum to at least $2s$ and $x_e\geq c\cdot\delta-s$.
In each case,~\cref{eq:matching-surplus-merge} gives $S(T,\mathcal{M})\geq s$.
All other components remain unchanged.\qedhere
\end{proof}

The matching-surplus bound now allows us to prove the main result for randomized matching, which we restate below.
\randomizedtheorem*
\begin{proof}
\Cref{alg:randomized} accepts only edges with two unmatched endpoints and never removes an accepted edge, so it maintains a valid online matching.
By~\cref{lem:marginals} and linearity of expectation,
\[
\E[|M_t|]=\sum_{e\in E(G_t)}x_e\text{.}
\]
For a component $T$ containing an edge, choose a maximum matching $\mathcal{M}$.
By~\cref{lem:randomized-certificate} and nonnegativity of the weights $y_v$,
\[
\sum_{e\in E(T)}x_e\geq c\cdot|\mathcal{M}|+s=c\cdot\opt(T)+s\text{.}
\]
Isolated vertices contribute zero to both the total weight and the optimum.
Summing over components and using $s\geq0$ gives $\E[|M_t|]\geq\frac47\opt_t$.\qedhere
\end{proof}

\section{Hardness for fractional matching in the edge arrival model}\label{sec:hardness}

We now proceed to show our hardness result. We show in this
section that
any $c$-competitive fractional online matching algorithm in the
edge arrival model satisfies $c \leq 4/7$.
We do so by constructing a family of growing trees of maximum degree four
that rules out every competitive ratio greater than $4/7$.

We move on to presenting
the construction.
The whole construction proceeds in $2L+2$ stages, where $L$ is the
parameter of our choice. As a result of applying the construction
for parameter $L$, we obtain a growing tree referred to as $T_L$.
The construction starts with an initialization
stage
(Stage $0$), then it proceeds through $2L$ intermediate
stages numbered from $1$ to $2L$, ending up with a completion
stage
(Stage $2L+1$). Each
edge $e$ that we ever add to the tree is presented at a specific
stage called $\mathrm{stage}(e)$. After
every two intermediate stages
a new set of
active vertices $A_\ell$ is exposed, used to proceed
with the construction for the next two consecutive stages.

In the initialization stage $0$, we present the first edge
$uv$ and we set the active set $A_0=\{u,v\}$, so both endpoints
of the newly presented edge become active. 

We proceed with the construction of the intermediate stages. 
For each $\ell \in \{1,\ldots,L \}$,
we define two consecutive stages: Stage $2\ell-1$ and Stage $2\ell$.
In Stage $2\ell-1$ we append a private pendant $w'$ to each
active vertex $w \in A_{\ell-1}$. In Stage $2\ell$, we append
a private dummy vertex $d_w$ to every active vertex
$w \in A_{\ell-1}$, and to each previously added pendant
$w'$ we append a private pendant $w''$.
Each pendant introduces a new edge to a vertex it is appended
to. 
Finally, we declare a new set of active vertices
$A_\ell = \bigcup_{w \in A_{\ell-1}} \{ w', w'' \}$, i.e.,
all the appended pendants but the dummy ones become active
for the next iteration. 

\begin{figure}[ht]
\centering
\begin{tikzpicture}[
  vtx/.style={circle,fill,inner sep=1.3pt},
  dummy/.style={circle,draw,fill=white,inner sep=1.3pt},
  ph/.style={inner sep=1pt,font=\scriptsize},
  act/.style={line width=1.2pt}]
\node[vtx] (u0) at (2.500,0.000) {};
\node[vtx] (v0) at (7.500,0.000) {};
\node[dummy] (n1) at (2.500,-0.700) {};
\node[vtx] (n2) at (1.250,-1.500) {};
\node[vtx] (n3) at (3.750,-1.500) {};
\node[dummy] (n4) at (1.250,-2.200) {};
\node[vtx] (n5) at (0.625,-3.000) {};
\node[vtx] (n6) at (1.875,-3.000) {};
\node[dummy] (n7) at (0.625,-3.700) {};
\node[dummy] (n8) at (1.875,-3.700) {};
\node[dummy] (n9) at (3.750,-2.200) {};
\node[vtx] (n10) at (3.125,-3.000) {};
\node[vtx] (n11) at (4.375,-3.000) {};
\node[dummy] (n12) at (3.125,-3.700) {};
\node[dummy] (n13) at (4.375,-3.700) {};
\node[dummy] (n14) at (7.500,-0.700) {};
\node[vtx] (n15) at (6.250,-1.500) {};
\node[vtx] (n16) at (8.750,-1.500) {};
\node[dummy] (n17) at (6.250,-2.200) {};
\node[vtx] (n18) at (5.625,-3.000) {};
\node[vtx] (n19) at (6.875,-3.000) {};
\node[dummy] (n20) at (5.625,-3.700) {};
\node[dummy] (n21) at (6.875,-3.700) {};
\node[dummy] (n22) at (8.750,-2.200) {};
\node[vtx] (n23) at (8.125,-3.000) {};
\node[vtx] (n24) at (9.375,-3.000) {};
\node[dummy] (n25) at (8.125,-3.700) {};
\node[dummy] (n26) at (9.375,-3.700) {};
\draw[act] (u0) -- node[ph,above]{0} (v0);
\draw (u0) -- node[ph,right=1pt]{2} (n1);
\draw[act] (n2) -- node[ph,above]{2} (n3);
\draw (n2) -- node[ph,right=1pt]{4} (n4);
\draw[act] (n5) -- node[ph,above]{4} (n6);
\draw (n5) -- node[ph,right=1pt]{5} (n7);
\draw (n6) -- node[ph,right=1pt]{5} (n8);
\draw (n2) -- node[ph,above left]{3} (n5);
\draw (n3) -- node[ph,right=1pt]{4} (n9);
\draw[act] (n10) -- node[ph,above]{4} (n11);
\draw (n10) -- node[ph,right=1pt]{5} (n12);
\draw (n11) -- node[ph,right=1pt]{5} (n13);
\draw (n3) -- node[ph,above left]{3} (n10);
\draw (u0) -- node[ph,above left]{1} (n2);
\draw (v0) -- node[ph,right=1pt]{2} (n14);
\draw[act] (n15) -- node[ph,above]{2} (n16);
\draw (n15) -- node[ph,right=1pt]{4} (n17);
\draw[act] (n18) -- node[ph,above]{4} (n19);
\draw (n18) -- node[ph,right=1pt]{5} (n20);
\draw (n19) -- node[ph,right=1pt]{5} (n21);
\draw (n15) -- node[ph,above left]{3} (n18);
\draw (n16) -- node[ph,right=1pt]{4} (n22);
\draw[act] (n23) -- node[ph,above]{4} (n24);
\draw (n23) -- node[ph,right=1pt]{5} (n25);
\draw (n24) -- node[ph,right=1pt]{5} (n26);
\draw (n16) -- node[ph,above left]{3} (n23);
\draw (v0) -- node[ph,above left]{1} (n15);
\node[above=2pt] at (u0) {$w$};
\node[above left=1pt] at (n2) {$w'$};
\node[above=2pt] at (n3) {$w''$};
\node[below=1pt] at (n1) {$d_w$};
\foreach \l in {0,1,2} \node[anchor=east] at (-0.6,-1.5*\l) {$A_{\l}$};
\end{tikzpicture}
\caption{The tree $T_2$ constructed for $L=2$.
Row $\ell$ shows the active vertices $A_\ell$, and the hollow vertices are dummies. Every edge is labeled by the number of the stage in which it is presented. The thick edges pair the vertices of each active set.}\label{fig:TL}
\end{figure}
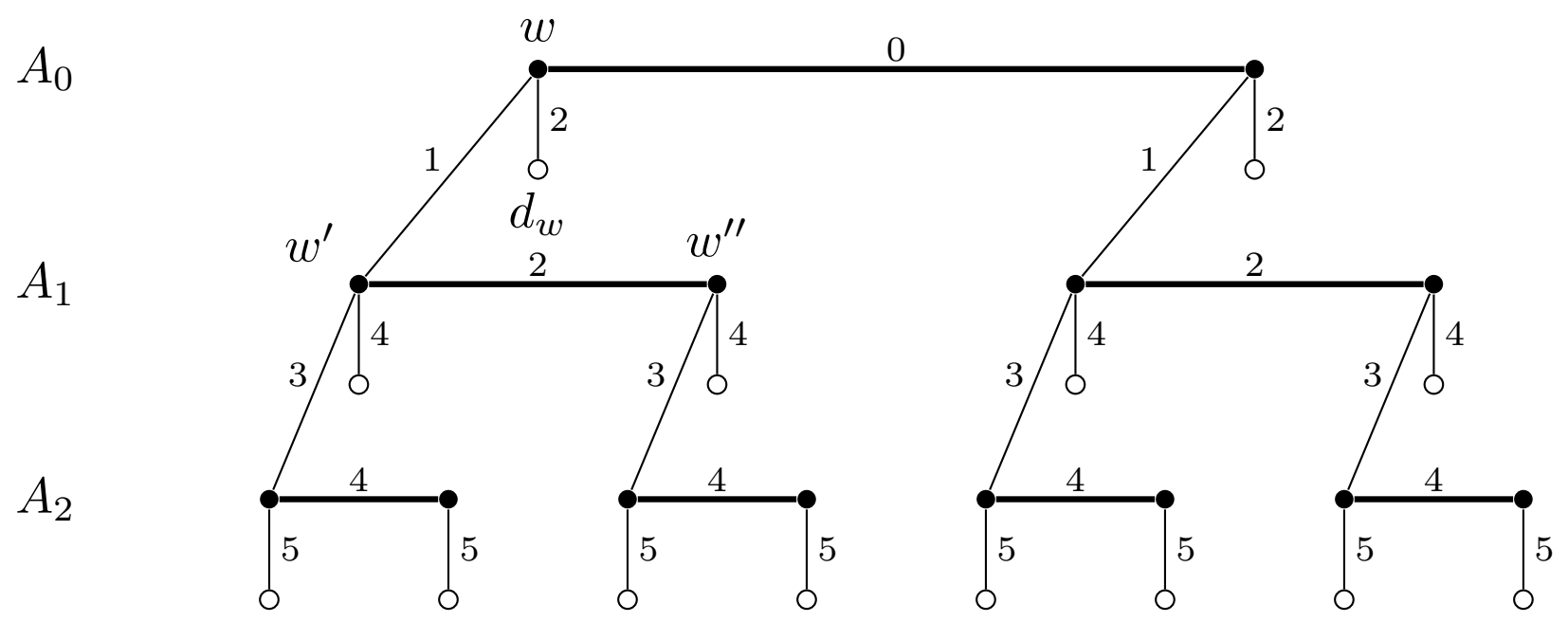

We conclude the construction
with the completion stage $2L+1$, where we append a dummy
vertex $d_w$ to every active vertex $w \in A_L$. This
construction is summarized in Table~\ref{tab:construction-stages}
and illustrated in Figure~\ref{fig:TL}.

Every edge after the first attaches a new leaf.
A vertex has degree at most two when it becomes active and subsequently receives at most two further incident edges.
Hence $T_L$ is a growing tree of maximum degree at most four.

\begin{table}[ht]
  \centering
  \setlength{\tabcolsep}{7pt}
  \renewcommand{\arraystretch}{1.2}
  \begin{tabularx}{\linewidth}{@{}l >{\raggedright\arraybackslash}X l@{}}
    \toprule
    stage $i$ & operations & active set after the stage \\
    \midrule
    $0$
      & $\create(u,v)$
      & $A_0 := \{u,v\}$ \\
    \addlinespace[4pt]
    $2\ell-1$
      & For every $w\in A_{\ell-1}$:\newline
        $\append(w,w')$
      &  \\
    \addlinespace[4pt]
    $2\ell$
      & For every $w\in A_{\ell-1}$:\newline
        $\append(w,d_w)$, then $\append(w',w'')$
      & $\displaystyle A_\ell := \bigcup_{w\in A_{\ell-1}}\{w',w''\}$ \\
    \addlinespace[4pt]
    $2L+1$
      & For every $w\in A_L$:\newline
        $\append(w,d_w)$
      &  \\
    \bottomrule
  \end{tabularx}

  \caption{Stages of the construction and the corresponding active
  sets. Method $\create(u,v)$ creates a new edge, while method
  $\append(u,v)$ appends a leaf $v$ to a tree vertex $u$.}
  \label{tab:construction-stages}
\end{table}

To analyze this construction,
we relate the matching formed by any algorithm to a vertex cover
that bounds its competitiveness from above.
To do so, we denote as $H_i$ the tree after stage~$i$ (for $i \in \{ 0,
\ldots, 2L+1 \}$),
i.e., $H_i$ is the subtree of $T_L$ consisting of edges $e \in E(T_L)$
with $\mathrm{stage}(e) \leq i$. Let $\alg_i$ and $\opt_i$ be the
value of the fractional matching algorithm and the size of a maximum matching of $H_i$.
We introduce a sequence of parameters $b_0 \geq b_1 \geq \ldots
\geq b_{2L+1}>b_{2L+2}=0$, to be chosen later, that we refer to as \emph{stage values}.
Write $S=2L+1$ and $\lambda_i=b_i-b_{i+1}$ for $i\in\{0,\ldots,S\}$.
We will assign a nonnegative cover weight $y_v$ to each vertex $v$.
We will maintain the invariant that for every stage $i$ and every
edge $uv$ in $H_i$ it holds that
\begin{equation}\label{eq:cover}
y_u+y_v\ \ge\ b_i\text{.}
\end{equation}
For every $i$ with $b_i>0$, the weights $y_v/b_i$ satisfy the
vertex-cover LP constraints from \cref{sec:algorithm} on $H_i$.
Our proof follows the LP-duality approach of Gamlath et al.~\cite[Section~2]{GamlathKMSW19}, combining matching feasibility with competitive guarantees at stage boundaries.
Balig\'acs et al.~\cite[Section~3.2]{BaligacsBDFGKPZ26} use a related argument for incremental vertex cover.

\begin{lemma}\label{lem:stage}
Under condition~\eqref{eq:cover}, any fractional matching algorithm satisfying
$\alg_i\ge c\cdot\opt_i$ after every stage satisfies
\begin{equation}\label{eq:ratio}
c\le\frac{\sum_{v\in V(T_L)}y_v}{\sum_{i=0}^{S}\lambda_i\opt_i}\text{.}
\end{equation}
\end{lemma}

\begin{proof}
Let $x_e$ be the weight assigned to edge $e$ by the
fractional matching algorithm.
Then, $x_e$ contributes to $\alg_i$ for every stage
$i\ge\mathrm{stage}(e)$.
Exchanging the order of summation and telescoping the coefficients
gives
\begin{align*}
\sum_{i=0}^S\lambda_i\alg_i
&=\sum_{e\in E(T_L)}x_e\sum_{i=\mathrm{stage}(e)}^S\lambda_i
=\sum_{e\in E(T_L)}b_{\mathrm{stage}(e)}\,x_e\\
&\le\sum_{uv\in E(T_L)}(y_u+y_v)\,x_{uv}
=\sum_{v \in V(T_L)}y_v\sum_{e\ni v}x_e
\le\sum_{v \in V(T_L)}y_v\text{.}
\end{align*}
Since $\lambda_i\ge0$, the guarantees after the stages give
\[
c\sum_{i=0}^S\lambda_i\opt_i\ \le\ \sum_{i=0}^S\lambda_i\alg_i\ \le\ \sum_{v\in V(T_L)}y_v\text{.}
\]
The denominator is positive because $\lambda_S=b_S>0$ and $\opt_S\ge1$.
Dividing by it gives \eqref{eq:ratio}.\qedhere
\end{proof}

\subsection{Assignment of stage values and cover weights}\label{sec:sigmaL}

To complete the argument, it remains to choose the stage values
$b_0,\ldots,b_{2L+1}$ and assign values $y_v$ appropriately.
Note that for $\ell \in \{ 0, \ldots, L \}$ it holds that
$|A_\ell|=2^{\ell+1}$, as we have $|A_0|=2$ and $|A_\ell|=2 \cdot |A_{\ell-1}|$.
For $\ell\in\{0,\ldots,L\}$, let $z_\ell=2^{-\ell}$.
Then $z_\ell|A_\ell|=2$, and each vertex in $A_\ell$ will receive weight $z_\ell$.

Let $b_0=2$, and let $y_v=y_u=z_0=1$
for the vertices $u,v$ connected via an edge in Stage $0$.
Clearly, condition~\eqref{eq:cover} is satisfied for $H_0$, which
consists of a single edge $uv$.
For each $\ell \in \{ 1, \ldots, L \}$, we set the stage values as
$b_{2\ell-1}=\frac{3}{2} z_{\ell-1}$ and $b_{2\ell}=z_{\ell-1}$,
and we also assign cover values to newly added pendants:
for every $w \in A_{\ell-1}$ we let
$y_{w'}=z_\ell=\frac{1}{2} z_{\ell-1}$,
$y_{w''}=z_\ell = \frac{1}{2} z_{\ell-1}$ and  $y_{d_w}=0$.
Finally, we let $b_{2L+1}=z_L$,
and for any vertex $d_w$ connected
to the tree in the completion stage we let $y_{d_w}=0$.
The stage values are displayed in Table~\ref{tab:stage-values}
in the middle column, and we present the endpoint sums for the
edges of the corresponding stage in the rightmost column.
Each edge $uv$ added in stage $i$ satisfies $y_u+y_v=b_i$.
The definitions above and $z_\ell=z_{\ell-1}/2$ show that the stage values are nonincreasing.
Thus an edge added in stage $j\leq i$ satisfies $y_u+y_v=b_j\geq b_i$, establishing~\eqref{eq:cover} for every edge of $H_i$.

\begin{table}[htbp]
\centering
\begin{tabular}{@{}lcr@{:\ }l@{}}
\toprule
stage $i$ & stage value $b_i$ & \multicolumn{2}{c@{}}{edge and endpoint sum} \\
\midrule
$0$ & $2$ & $uv$ & $1+1$ \\
$2\ell-1$ & $\frac32 z_{\ell-1}$ & $ww'$ & $z_{\ell-1}+\frac12 z_{\ell-1}$ \\
\multirow{2}{*}{$2\ell$} & \multirow{2}{*}{$z_{\ell-1}$} & $wd_w$ & $z_{\ell-1}+0$ \\
 & & $w'w''$ & $\frac12 z_{\ell-1}+\frac12 z_{\ell-1}$ \\
$2L+1$ & $z_L$ & $wd_w$ & $z_L+0$ \\
\bottomrule
\end{tabular}
\caption{Stage values and endpoint sums.}
\label{tab:stage-values}
\end{table}

To use Lemma~\ref{lem:stage}, we now need to bound the term
$\frac{\sum_{v\in V(T_L)}y_v}{\sum_{i=0}^S\lambda_i\opt_i}$.
In order to compute the denominator,
we let $\Delta_i=\opt_i-\opt_{i-1}$ be the increase in the
optimum matching size in stage
$i \in \{ 0, \ldots, 2L+1 \}$, with $\opt_{-1}=0$.
Since $\lambda_i=b_i-b_{i+1}$ and $b_{2L+2}=0$, regrouping the sum
by $\opt_i$ gives
$\sum_{i=0}^{2L+1} \lambda_i\opt_i=\sum_{i=0}^{2L+1} b_i\Delta_i$,
which
we compute stage by
stage.

At the end of every stage, $H_i$ has a perfect matching.
We maintain one that, after every even stage $2\ell$, contains a perfect matching on $A_\ell$.
Initially this is the edge $uv$.
In stage $2\ell-1$, replace each matching edge on $A_{\ell-1}$ by the two edges to its endpoints' new leaves.
In stage $2\ell$, replace each $ww'$ by $wd_w$ and $w'w''$.
In the completion stage, replace each matching edge on $A_L$ by the two edges to its endpoints' dummy leaves.

Thus $\opt_i=|V(H_i)|/2$.
Counting the new vertices gives
\[
\Delta_0=1,\qquad
\Delta_{2\ell-1}=\frac{|A_{\ell-1}|}{2},\qquad
\Delta_{2\ell}=|A_{\ell-1}|,\qquad
\Delta_{2L+1}=\frac{|A_L|}{2}\text{.}
\]
Multiplying by the corresponding stage values gives $2$, $\frac32$, $2$, and $1$, respectively, as in Table~\ref{tab:increases}.

Each intermediate stage $i\in\{2\ell-1,2\ell\}$ adds $|A_{\ell-1}|$ vertices of weight $z_\ell$.
The additional dummy vertices in stage $2\ell$ have weight zero, so each intermediate stage increases the total cover weight by $|A_{\ell-1}|z_\ell=1$.
These increases are listed in the third column of Table~\ref{tab:increases}.

\begin{table}
 \begin{center}
\medskip
\begin{tabular}{@{}lcc@{}}
\toprule
stage $i$ &  $b_i\Delta_i$ & increase in $\sum_v y_v$\\
\midrule
$0$ &  $2$ & $2$\\
$2\ell-1$  & $\frac32$ & $1$\\
$2\ell$  &  $2$ & $1$\\
$2L+1$ & $1$ & $0$\\
\bottomrule
\end{tabular}
\end{center}
\caption{Values $b_i \Delta_i$ and the increase in
$\sum_v y_v$ in each stage $i$.}
\label{tab:increases}
\end{table}

We are now ready to restate and prove our hardness result.
\hardnesstheorem*
\begin{proof}
Consider a $c$-competitive fractional matching algorithm in the edge arrival model,
run on $T_L$ with edges presented in the order of the construction stages.
By Lemma~\ref{lem:stage}, it holds that
$c \leq \frac{\sum_{v\in V(T_L)}y_v}{\sum_{i=0}^{2L+1}\lambda_i\opt_i}$.
From Table~\ref{tab:increases}, we get $\sum_{v\in V(T_L)}y_v = 2+2L$
and
$\sum_{i=0}^{2L+1}\lambda_i\opt_i=3+\frac{7}{2}L$.
Thus $c \leq \frac{2+2L}{3+\frac{7}{2}L}=\frac{4(L+1)}{7L+6}$.
For every $c>4/7$, choose $L$ sufficiently large that $\frac{4(L+1)}{7L+6}<c$.
Hence no fractional matching algorithm can satisfy the competitive guarantee at every prefix of this fixed input.
For a randomized algorithm, its expected edge weights form a fractional matching, so the same argument applies.\qedhere

\end{proof}

\subparagraph*{Incremental matching.}
The hardness bound also holds when the entire edge arrival sequence is known in advance, provided that matching decisions remain irrevocable and the competitive guarantee is required after every prefix.
Indeed, the proof of \cref{lem:stage} uses only the feasibility of the fixed edge weights, not how they were chosen.
Together with \cref{thm:fractional,thm:randomized}, this gives the optimal ratio $4/7$ for fractional and randomized matching on forests in the analogue of the incremental model studied by Balig\'{a}cs et al.~\cite{BaligacsBDFGKPZ26}.
For comparison, they prove an optimal ratio of $3/2$ for incremental fractional vertex cover; offline matching--cover duality does not imply reciprocal competitive ratios for these evolving solutions.

\subparagraph*{AI methodology.}\label{sec:ai-usage-summary}
We used AI agents to assist with the research and preparation of this paper,
primarily models from the ChatGPT and Claude suites, with occasional use of DeepSeek 3.1-Flash and Xiaomi MiMo-v2.6-pro.
Notably, the results described in Sections \ref{sec:algorithm}, \ref{sec:randomized}, and \ref{sec:hardness} were discovered
by AI agents after significant guidance and refined by the authors to be as simple and clear as possible.
The authors have verified the correctness and originality of the results and the accuracy of the references.

\bibliography{paper}

@inproceedings{KarpVV90,
  author    = {Karp, Richard M. and
               Vazirani, Umesh V. and
               Vazirani, Vijay V.},
  title     = {An Optimal Algorithm for On-line Bipartite Matching},
  year      = {1990},
  booktitle = {{STOC} 1990: 22nd Annual {ACM} Symposium on Theory of Computing},
  pages     = {352--358},
  doi       = {10.1145/100216.100262},
}

@article{Birnbaum,
author = {Birnbaum, Benjamin and Mathieu, Claire},
title = {On-line bipartite matching made simple},
year = {2008},
issue_date = {March 2008},
publisher = {Association for Computing Machinery},
address = {New York, NY, USA},
volume = {39},
number = {1},
issn = {0163-5700},
url = {https://doi.org/10.1145/1360443.1360462},
doi = {10.1145/1360443.1360462},
journal = {SIGACT News},
month = mar,
pages = {80–87},
numpages = {8}
}

@inproceedings{BuchbinderJN07,
  author    = {Buchbinder, Niv and
               Jain, Kamal and
               Naor, Joseph (Seffi)},
  title     = {Online Primal-Dual Algorithms for Maximizing Ad-Auctions Revenue},
  year      = {2007},
  booktitle = {{ESA} 2007: 15th Annual European Symposium on Algorithms},
  series    = {LNCS},
  volume    = {4698},
  pages     = {253--264},
  doi       = {10.1007/978-3-540-75520-3_24},
}

@inproceedings{DevanurJK13,
  author    = {Devanur, Nikhil R. and
               Jain, Kamal and
               Kleinberg, Robert D.},
  title     = {Randomized Primal-Dual Analysis of {RANKING} for Online Bipartite Matching},
  year      = {2013},
  booktitle = {{SODA} 2013: 24th Annual {ACM}-{SIAM} Symposium on Discrete Algorithms},
  pages     = {101--107},
  doi       = {10.1137/1.9781611973105.7},
}

@inproceedings{WangW15,
  author    = {Wang, Yajun and
               Wong, Sam Chiu-wai},
  title     = {Two-sided Online Bipartite Matching and Vertex Cover: Beating the Greedy Algorithm},
  year      = {2015},
  booktitle = {{ICALP} 2015: 42nd International Colloquium on Automata, Languages, and Programming},
  series    = {LNCS},
  volume    = {9134},
  pages     = {1070--1081},
  doi       = {10.1007/978-3-662-47672-7_87},
  archivePrefix = {arXiv},
  eprint    = {1305.1694},
}

@inproceedings{BoyarFKL17,
  author    = {Boyar, Joan and
               Favrholdt, Lene M. and
               Kotrb{\v c}{\'i}k, Michal and
               Larsen, Kim S.},
  title     = {Relaxing the Irrevocability Requirement for Online Graph Algorithms},
  year      = {2017},
  booktitle = {{WADS} 2017: 15th International Symposium on Algorithms and Data Structures},
  series    = {Lecture Notes in Computer Science},
  volume    = {10389},
  pages     = {217--228},
  doi       = {10.1007/978-3-319-62127-2_19},
}

@article{Boyar,
author = {Boyar, Joan and Favrholdt, Lene M. and Kotrb{\v c}{\'i}k, Michal and Larsen, Kim S.},
title = {Relaxing the Irrevocability Requirement for Online Graph Algorithms},
year = {2022},
issue_date = {Jul 2022},
publisher = {Springer-Verlag},
address = {Berlin, Heidelberg},
volume = {84},
number = {7},
issn = {0178-4617},
url = {https://doi.org/10.1007/s00453-022-00944-w},
doi = {10.1007/s00453-022-00944-w},
journal = {Algorithmica},
month = jul,
pages = {1916–1951},
numpages = {36}
}

@inproceedings{BuchbinderST17,
  author    = {Buchbinder, Niv and
               Segev, Danny and
               Tkach, Yevgeny},
  title     = {Online Algorithms for Maximum Cardinality Matching with Edge Arrivals},
  year      = {2017},
  booktitle = {{ESA} 2017: 25th Annual European Symposium on Algorithms},
  series    = {LIPIcs},
  volume    = {87},
  pages     = {22:1--22:14},
  doi       = {10.4230/LIPIcs.ESA.2017.22},
}

@article{BuchbinderST19,
  author    = {Buchbinder, Niv and
               Segev, Danny and
               Tkach, Yevgeny},
  title     = {Online Algorithms for Maximum Cardinality Matching with Edge Arrivals},
  year      = {2019},
  journal   = {Algorithmica},
  volume    = {81},
  number    = {5},
  pages     = {1781--1799},
  doi       = {10.1007/s00453-018-0505-7},
}

@inproceedings{GamlathKMSW19,
  author    = {Gamlath, Buddhima and
               Kapralov, Michael and
               Maggiori, Andreas and
               Svensson, Ola and
               Wajc, David},
  title     = {Online Matching with General Arrivals},
  year      = {2019},
  booktitle = {{FOCS} 2019: 60th Annual {IEEE} Symposium on Foundations of Computer Science},
  pages     = {26--37},
  doi       = {10.1109/FOCS.2019.00011},
  archivePrefix = {arXiv},
  eprint    = {1904.08255},
}

@inproceedings{BuchbinderNW23,
  author    = {Buchbinder, Niv and
               Naor, Joseph (Seffi) and
               Wajc, David},
  title     = {Lossless Online Rounding for Online Bipartite Matching (Despite its Impossibility)},
  year      = {2023},
  booktitle = {{SODA} 2023: 34th Annual {ACM}-{SIAM} Symposium on Discrete Algorithms},
  pages     = {2030--2068},
  doi       = {10.1137/1.9781611977554.ch78},
  archivePrefix = {arXiv},
  eprint    = {2106.04863},
}

@inproceedings{PashkovichS25,
  author    = {Pashkovich, Kanstantsin and
               Snow, Thomas},
  title     = {Online Algorithm for Fractional Matchings with Edge Arrivals in Graphs of Maximum Degree Three},
  year      = {2025},
  booktitle = {{WAOA} 2025: 23rd International Workshop on Approximation and Online Algorithms},
  series    = {LNCS},
  volume    = {16077},
  pages     = {198--212},
  doi       = {10.1007/978-3-032-06706-7_14},
  archivePrefix = {arXiv},
  eprint    = {2602.07355},
}

@inproceedings{BaligacsBDFGKPZ26,
  author    = {Balig{\'a}cs, J{\'u}lia and
               Bosek, Bart{\l}omiej and
               Disser, Yann and
               Feldmann, Andreas Emil and
               Gutowski, Grzegorz and
               K{\k{e}}pi{\'n}ska, Katarzyna and
               Putra, Pawe{\l} and
               Zych-Pawlewicz, Anna},
  title     = {Online and Incremental Fractional Vertex Cover on Trees},
  year      = {2026},
  booktitle = {{ESA} 2026: 34th Annual European Symposium on Algorithms},
  series    = {LIPIcs},
  volume    = {388},
  pages     = {158:1--158:15},
  doi       = {10.4230/LIPIcs.ESA.2026.158},
  archivePrefix = {arXiv},
  eprint    = {2608.24630},
}

@mastersthesis{Bromblik26,
  author    = {Bromblik, Bartosz},
  title     = {Fractional Vertex Cover Online on Trees},
  year      = {2026},
  school    = {Jagiellonian University},
  type      = {{BSc} thesis},
  note      = {Analytical Computer Science},
}

@unpublished{Bienkowski26,
  author    = {Bienkowski, Marcin},
  title     = {A 1.764-competitive algorithm for online fractional vertex cover on forests},
  year      = {2026},
  note      = {Unpublished manuscript},
}

@inproceedings{JiangZ26,
  author    = {Jiang, Tianle and
               Zhang, Yuhao},
  title     = {Edge Arrival Online Matching: The Power of Free Disposal on Acyclic Graphs},
  year      = {2026},
  booktitle = {{WINE} 2024: 20th International Conference on Web and Internet Economics},
  series    = {LNCS},
  volume    = {15534},
  pages     = {591--608},
  doi       = {10.1007/978-3-032-08560-3_33},
  archivePrefix = {arXiv},
  eprint    = {2410.21727},
}

@misc{Lu26,
  author    = {Lu, Tianhang},
  title     = {A New Lower Bound for Online Vertex Cover under Vertex Arrivals},
  year      = {2026},
  pages     = {1--12},
  archivePrefix = {arXiv},
  eprint    = {2608.09210},
}

@misc{Tang26,
  author    = {Tang, Zhihao Gavin},
  title     = {Optimal Competitive Ratio of Two-sided Online Bipartite Matching},
  year      = {2026},
  pages     = {1--16},
  archivePrefix = {arXiv},
  eprint    = {2602.18049},
}

@misc{TangZ26,
  author    = {Tang, Zhihao Gavin and
               Zhang, Yuhao},
  title     = {A Tight Bound on Online Vertex Cover under Edge Arrivals},
  year      = {2026},
  pages     = {1--9},
  archivePrefix = {arXiv},
  eprint    = {2608.04994},
}

\end{document}